\documentclass[12pt]{article} \def\theequation{\arabic{section}.\arabic{equation}}
\newcounter{rown}

\usepackage[centertags]{amsmath}
\usepackage{color}
\usepackage{amsfonts}
\usepackage{amssymb}
\usepackage{amsthm}
\usepackage{newlfont}
\usepackage{amsthm}

\begin{document}
\renewcommand{\theequation}{\arabic{section}.\arabic{equation}}
\title{{\bf Minimal $D$=4 supergravity from the superMaxwell algebra
\vskip 1.5cm}}
\author{ J.A. de Azc\'{a}rraga, \\
Departamento de F\'{\i}sica Te\'orica and IFIC (CSIC-UVEG), \\
 46100-Burjassot (Valencia), Spain\\
 J. M. Izquierdo,\\
Departamento de F\'{\i}sica Te\'orica, Universidad de Valladolid, \\
47011-Valladolid, Spain}
\date{\small{March 17, 2014}}
\maketitle
\vskip 1cm

\begin{abstract}
We show that the  first-order  $D$=4, $N=1$ pure supergravity lagrangian
four-form can be obtained geometrically as a quadratic expression
in the curvatures of the Maxwell superalgebra. This is achieved
by noticing that the relative coefficient between the two
terms of the Lagrangian that makes the action locally supersymmetric
also determines trivial field equations for the gauge fields
associated with the extra generators of the Maxwell superalgebra.
Along the way, a convenient geometric procedure to check the local
supersymmetry of a class of lagrangians is developed.
\end{abstract}
\newpage

\section{Introduction}
Since the advent of supersymmetry, there has been
an interest in superalgebras going beyond the standard
superPoincar\'e one. For instance, enlarged $D$=11 supersymmetry algebras
were considered by D'Auria and Fr\'e in \cite{DF82} and further in
\cite{BAIPV04} in a search for the group structure underlying $D=11$
supergravity \cite{CJS-78}, which is hidden due to the presence of the
three-form that needs being trivialized as a product of one-forms
to be associated with Maurer-Cartan (MC) forms.
The resulting superalgebras go beyond the $D$=11 superPoincar\'e
algebra and contain additional fermionic generators and tensorial charges.
Larger supersymmetry algebras (and correspondingly enlarged
superspaces), also appear associated with
super-$p$-branes, where the tensorial charges are realized
as topological charges \cite{AGIT89} (see \cite{TS97}
for the M5-brane and further \cite{CAIP00}). The 560-dimensional
$D$=11 superlagebra includes 528 vector and tensorial charges and is
usually referred to as the M-theory
superalgebra \cite{T97}; fermionic extensions of the superPoincar\'e
algebra, similar to those of  D'Auria and Fr\'e but without tensorial
charges, were introduced by Green in \cite{G89} by adding an
extra spinorial generator. The Green algebra was used by
Siegel \cite{Siegel-94} to produce a superstring action with a
manifestly supersymmetric Wess-Zumino term, a procedure further
generalized by Bergshoeff and Sezgin to super-$p$-branes by introducing
larger Green-type superalgebras \cite{BS95} (see also \cite{S97}).
These algebras can be viewed as the result of
successive extensions of the supertranslations algebra \cite{CAIP00},
and the associated enlarged superspace group manifolds may
be used to construct strictly invariant (rather than invariant up to
total derivative) Wess-Zumino terms for general $p$-branes,
as discussed in detail in \cite{CAIP00}.

  In a separate context, Hatsuda and Sakaguchi showed that there
is a suitable flat limit for the $AdS$ superstring that leads to bilinear
WZ terms and to enlarged Poincar\'e superalgebras \cite{HS02}.
They interpreted these \cite{HS03} as what now are
termed (super)algebra expansions, which were
studied in detail in \cite{AIPV03} (see also \cite{IRS07}
for an expansion-related procedure). Expansions
are obtained from the original algebras by means of
a series expansion (hence their name \cite{AIPV03}) of
their MC one-forms . As a result, the
expansion procedure leads to algebras of higher dimension than
the original one; nevertheless, the dimension preserving
\.In\"on\"u-Wigner contraction (and its Weimar-Woods
generalization \cite{W-W-95}) are obtained as particular
cases \cite{AIPV03}. Expansions were shown to lead to the full ({\it i.e.}
including the $D$=11 Lorentz algebra) M-theory superalgebra, which is a
particular expansion of $osp(1|32)$ \cite{AIPV03}.
The ($p,q$)-superPoincar\'e algebras \cite{AP-89} governing
the $D$=3 extended supergravities have also been shown \cite{AzIz-11}
to be related to particular expansions of $osp(p+q|2)$.

  Recently, a $D$=4 Maxwell superalgebra has been introduced
in \cite{BGKL10} as the minimal superalgebra that contains the
Maxwell algebra as its bosonic subalgebra (see \cite{BCR70} for
Maxwell algebras). This Maxwell superalgebra
can be viewed as an extension of the $D$=4 Green algebra by the
tensorial charges algebra, and it was used in \cite{BGKL10} to construct
a superparticle model. But the Maxwell algebra is also
an expansion of the $D$=4 $adS$ algebra $o(3,2)$ \cite{AILW-12}.
The minimal Maxwell superalgebra $s\mathcal{M}$
to be considered here also follows from an expansion
of the $D$=4 $adS$ superalgebra $osp(N|4)$ (further
$N$-extended superMaxwell algebras are also described
in \cite{AILW-12} using the expansion method).

The geometric approach to supergravity has a long history (see {\it e.g.}
\cite{CdAF} and references therein).
In this  paper we wish to  show that the minimal $D$=4 Maxwell superalgebra
$s\mathcal{M}$ may be used to  derive the action of the basic (or minimal) $N$=1, $D$=4
supergravity from the curvature forms of the Cartan structure equations
associated with $s\mathcal{M}$. This extends to the supersymmetric case
the $D$=4 gravity results obtained from the bosonic Maxwell
algebra  \cite{AKL11} (the $D$=1+2 gravity
case has been considered very recently in \cite{Ho-Re-14}).
To this aim, we first briefly review the Maxwell algebra
and its relation to ordinary gravity. In Sec.~3, a
family of lagrangian forms depending on a parameter $k$ will be constructed
geometrically in terms of curvatures associated with the $D$=4 Maxwell
superalgebra $s\mathcal{M}$. To show that a value of $k$
provides the lagrangian of $D$=4 minimal supergravity we
present first in Sec.~4, for a generic algebra, a procedure to analyze
the local invariance of a class of $D$=4 lagrangians
that includes the Chern-Simons lagrangians as a particular case.
Sec.~5  applies the method to the $s\mathcal{M}$ superalgebra
and to supergravity. A final section contains some comments.

 \section{Maxwell algebra and the gravity action}
 The $D$=4 Maxwell algebra ${\mathcal M}$ is given by the following commutators:
 \begin{eqnarray}
 \label{Maxcom}
 \left[ M_{ab}, M_{cd} \right] &=& \eta_{bc} M_{ad} - \eta_{ac} M_{bd}
 -\eta_{bd} M_{ac} + \eta_{ad} M_{bc}\ ,\nonumber\\
\left[ M_{ab},P_c \right] &=& \eta_{bc} P_a - \eta_{ac} P_b \ ,
\nonumber \\
\left[ P_a, P_b \right] &=& Z_{ab}\ ,
\nonumber\\
\left[ M_{ab}, Z_{cd} \right] &=& \eta_{bc} Z_{ad} - \eta_{ac}
Z_{bd} -\eta_{bd} Z_{ac} + \eta_{ad} Z_{bc}
 \end{eqnarray}
($a=0,\dots ,3$), where $\eta_{ab}$ is the (mostly plus) Minkowski metric.
Besides the Poincar\'e generators, the Maxwell algebra contains six
additional tensorial charges $Z_{ab}$ that extend centrally the abelian translation
algebra and that behave tensorially under the Lorentz algebra ${\mathcal L}$.

It is convenient to describe this algebra through its
MC eqs.~satisfied by the one-forms $\omega^{ab}$, $e^a$,
 $f^{ab}$ dual to the generators $M_{ab}$, $P_a$, $Z_{ab}$,
 \begin{equation}
 \label{duality}
   \omega^{ab}(M_{cd})= \delta_{cd}^{ab}\quad ,
   \quad e^a(P_b) = \delta^a_b \quad ,\quad
   f^{ab}(Z_{cd})= \delta_{cd}^{ab}\; .
\end{equation}
They are given by
\begin{eqnarray}\label{MCMax}
   0 &=& d\omega^{ab} + \omega^a{}_c \wedge \omega^{cb}\
   ,\nonumber\\
   0 &=& de^a  + \omega^a{}_b \wedge e^b \ ,\nonumber\\
   0 &=& df^{ab} + e^a\wedge e^b +\omega^a{}_c \wedge f^{cb} -
   \omega^b{}_c \wedge f^{ca} \ .
\end{eqnarray}

The `soft' version of these MC equations
introduce the gauge curvatures $R^{ab}, T^a$ and $F^{ab}$
in terms of the gauge field forms. Using without risk of confusion
the same notation for the MC one-forms and the gauge field ones,
the Cartan structure equations
$\Omega =d \theta + \theta \wedge \theta=
d \theta + \frac{1}{2}[\theta,\theta]$ where
$\theta= e^a P_a+  \frac{1}{2}\omega^{ab} M_{ab} + \frac{1}{2} f^{ab} Z_{ab}$
determine the various curvatures.
Writing $\Omega=\frac{1}{2} R^{ab}M_{ab}+T^a P_a +\frac{1}{2} F^{ab} Z_{ab}$,
they are found to be
\begin{eqnarray}
\label{CurvMax}
   R^{ab} &=& d\omega^{ab} + \omega^a{}_c \wedge \omega^{cb}\
   ,\nonumber\\
   T^a &=& de^a + \omega^a{}_b \wedge e^b \ ,\nonumber\\
   F^{ab} &=& df^{ab} + e^a\wedge e^b +\omega^a{}_c \wedge f^{cb}
   -\omega^b{}_c \wedge f^{ca}\ .
\end{eqnarray}
The Lorentz covariant differentials of the curvatures
$DR=dR+[\omega,R],\, DT= dT+[\omega,T]=  [R, e],\, DF=dF+[\omega,F]= [R,f]+[T,e]$
are then
\begin{eqnarray}
\label{DCurMax}
  DR^{ab} &=& (dR+\omega\wedge R - R\wedge \omega)^{ab} = 0\ ,\nonumber\\
  DT^a &=& R^{ab}\wedge e_b \ ,\nonumber\\
  DF^{ab} &=& R^a{}_c\wedge f^{cb}- R^b{}_c\wedge f^{ca} +
  T^a\wedge e^b - e^a\wedge T^b\ .
\end{eqnarray}

The following Lorentz invariant lagrangian four-form constructed
from the Maxwell curvatures (with length dimensions of an action in
$D$=4 ($L^2$) in geometrized $\kappa=1=c$) units,
was considered in \cite{AKL11}
\begin{equation}\label{4formMax}
    {\cal L} \sim \epsilon_{abcd} R^{ab}\wedge F^{cd} \ ;
\end{equation}
other possibilities were also discussed there. Since
the extra field $f_{ab}$ appears in an exterior
differential this lagrangian leads, up to boundary terms
that will be disregarded here, to the standard
Einstein-Hilbert action for gravity,
\begin{equation}
\label{Actiongrav}
    \int_M  {\cal L}_{EH} \sim \int_M
    \epsilon_{abcd} R^{ab} \wedge e^c\wedge e^d \; .
\end{equation}
Thus, since the gauging of the Maxwell group provides
a geometric framework to derive the gravity lagrangian,
it is natural to ask \cite{AKL11} whether  a
minimal supersymmetrization of the Maxwell algebra
may lead to the pure gravity lagrangian. Our aim
is to show that this is the case.

 \section{Maxwell superalgebra and geometric ingredients
 of minimal supergravity}
Pure, simple $D$=4 supergravity just includes
the graviton and the gravitino, with two on-shell
degrees of freedom each. To express its lagrangian
in terms of curvature bilinears
we consider the 24-dimensional minimal superMaxwell algebra
$s{\mathcal M}$ \cite{BGKL10}. It contains
the 16-dimensional Maxwell algebra \eqref{Maxcom} as its
even subalgebra, and the brackets involving the odd
generators are
\begin{eqnarray}
\label{sMaxCom}
    \left[ M_{ab} ,Q_\alpha\right] &=& \frac{1}{2}
    \gamma_{ab}{}^\beta{}_\alpha Q_\beta \; ,
    \nonumber\\
    \left[ M_{ab} ,\Sigma_\alpha\right] &=& \frac{1}{2}
    \gamma_{ab}{}^\beta{}_\alpha \Sigma_\beta \; ,
    \nonumber\\
    \left\{ Q_\alpha, Q_\beta\right\} &=&
    \gamma^a{}_{\alpha\beta} P_a \; ,
    \nonumber\\
\left[ P_a ,Q_\alpha\right] &=&\frac{1}{2}
\gamma_a{}^\alpha{}_\beta \Sigma_\beta \; ,
\nonumber\\
 \left\{ Q_\alpha, \Sigma_\beta\right\} &=& -\frac{1}{2}
 \gamma^{ab}{}_{\alpha\beta} Z_{ab}\ ,
\end{eqnarray}
where $Q_\alpha, \alpha=1,\dots 4$, is the
supersymmetry generator ($[Q]=L^{-1/2}$) and, as in the Green algebra,
the $[P,Q]$ commutator produces an additional spinor generator
$\Sigma_\alpha$,  $[\Sigma]=L^{-3/2}$. All spinors
above and below are Majorana spinors.

The dual MC one-forms of $s{\mathcal M}$ are defined by the
duality conditions
\eqref{duality} plus
\begin{equation}\label{Moreduality}
    \psi^\alpha(Q_\beta) = \delta^\alpha_\beta =
    \xi^\alpha(\Sigma_\beta)\; ,
\end{equation}
and their MC eqs., $0 = d \theta +\theta
\wedge \theta$, where now
$\theta= e^a P_a+ \frac{1}{2}\omega^{ab} M_{ab} + \frac{1}{2}
f^{ab} Z_{ab}+ \psi^\alpha  Q_\alpha + \xi^\alpha \Sigma_\alpha$,
are given by
\begin{eqnarray}
\label{MCmaxwelld4}
 0&=& d\omega^{ab} + \omega^a{}_c\wedge \omega^{cb}\ , \nonumber\\
 0&=& de^a + \omega^a{}_b\wedge e^b  -\frac{1}{2} {\bar \psi} \gamma_a \wedge \psi
  \ , \nonumber\\
  0&=&  df^{ab} + \omega^a{}_c\wedge f^{cb} - \omega^b{}_c\wedge f^{ca}
  + {\bar \xi} \gamma^{ab} \wedge \psi + e^a\wedge e^b \ ,\nonumber\\
0&=& d\psi + \frac{1}{4} \omega_{ab} \gamma^{ab} \wedge \psi \ , \nonumber\\
0&=& d \xi + \frac{1}{4} \omega_{ab}\gamma^{ab}\wedge \xi
   + \frac{1}{2} e_a\gamma^a \wedge \psi  \; .
\end{eqnarray}
We use $(\lambda \lambda{}')^*= \lambda^* \lambda{}'^*$ for the conjugation
of bilinears of odd scalars, so that both $\omega_{ab}$ and $e_a$ are real.
The gauge curvatures, again using the same notation for the gauge field
one-forms and for the MC ones, are given by the structure equations
$\Omega =d\theta + \theta \wedge \theta$, where
$\Omega = \frac{1}{2} R^{ab}M_{ab}+T^a P_a
 +\frac{1}{2} F^{ab} Z_{ab} + \Psi^\alpha Q_\alpha + \xi^\alpha\Sigma_\alpha$.
 Explicitly,
\begin{eqnarray}
\label{maxwelld4}
   R^{ab} &=& d\omega^{ab} + \omega^a{}_c\wedge \omega^{cb}\ , \nonumber\\
  T^a &=& de^a + \omega^a{}_b\wedge e^b-\frac{1}{2} {\bar \psi} \gamma_a \wedge \psi
  \ , \nonumber\\
  F^{ab} &=& df^{ab} + \omega^a{}_c\wedge f^{cb} - \omega^b{}_c\wedge f^{ca}
  +{\bar \xi} \gamma^{ab} \wedge \psi  + e^a\wedge e^b \nonumber\\
   \Psi^\alpha &=& d\psi^\alpha + \frac{1}{4} \omega_{ab} (\gamma^{ab} \wedge \psi)^\alpha \,
   \nonumber \\
     \Xi^\alpha &=& d \xi^\alpha + \frac{1}{4} \omega_{ab}(\gamma^{ab}\wedge \xi)^\alpha
   + \frac{1}{2} e_a(\gamma^a \wedge \psi)^\alpha  \; .
\end{eqnarray}
These curvatures have dimensions
$[R]=L^0,\, [\Psi]=L^{1/2},\, [T]=L,\, [\Xi]=L^{3/2}, [F]=L^2$.

 The Lorentz covariant exterior differential of the
curvatures is given by
$DR=dR+[\omega,R]=0,\, DT= dT+[\omega,T]=  [R, e] + [\Psi,
\psi],\, DF=dF+[\omega,F]= [R,f]+[T,e]+ [\Psi, \xi]
+ [\Xi, \psi]$, plus $D\Psi = d\Psi + [\omega,\psi]$ and
$D\Xi = d\Xi + [\omega,\Xi]+ [T, \psi] + [\Psi, e]$. Explicitly,
\begin{eqnarray}
\label{DFsuperM}
   DR^{ab} &=& (dR+\omega\wedge R - R\wedge \omega)^{ab} = 0\ ,\nonumber\\
  DT^a &=& R^{ab}\wedge e_b+ \bar{\psi} \gamma_a \wedge\Psi \ ,\nonumber\\
  DF^{ab} &=& R^a{}_c\wedge f^{cb}- R^b{}_c\wedge f^{ca} +
  T^a\wedge e^b - e^a\wedge T^b+ \bar{\Xi} \gamma^{ab} \wedge \psi-
 \bar{\xi} \gamma^{ab} \wedge \Psi\ , \nonumber\\
D\Psi^\alpha &=&  \frac{1}{4} (R_{ab} \gamma^{ab}\wedge \psi)^\alpha\,\nonumber
\\
D\Xi^\alpha &=&  \frac{1}{4}( R_{ab} \gamma^{ab}\wedge \xi)^\alpha  +
\frac{1}{2} T_a (\gamma^a\wedge \psi)^\alpha - \frac{1}{2} e_a (\gamma^a
\wedge \Psi)^\alpha\ .
\end{eqnarray}

To  show that $D$=4 minimal supergravity can be written in terms of
the above curvatures, consider Lagrangian four-forms $B$, $[B]=L^2$
given by linear combinations of the type
\begin{equation}
\label{maxcur}
    B = \epsilon_{abcd} R^{ab} \wedge F^{cd}
    + k\ {\bar \Xi} \gamma_5 \wedge \Psi\ ,
\end{equation}
where $k$ is a constant to be determined and
$\gamma_5=-\gamma^0\gamma^1\gamma^2\gamma^3, \gamma_5^2=-1$,
$\gamma_5 \gamma^{abcd}=\epsilon^{abcd}$ with
$\gamma^{abcd}=1/4! \gamma^{[a} \gamma^{b}\gamma^{c}\gamma^{d]}$.
It will turn out that there is a value of $k$ for which the field equations
for $f^{ab}$ and $\xi^\alpha$ are trivial (0=0). For this
value of $k$, the resulting action becomes the well-known action
of minimal $D$=4 supergravity, given by eq.~\eqref{Actiongrav}
plus the Rarita-Schwinger (R-S) action for the lagrangian
\begin{equation}
\label{acRS}
{\mathcal L}_{RS}\sim  {\bar \psi} \wedge \gamma_5 \gamma_a e^a \wedge D \psi
= \frac{1}{3!}\epsilon_{abcd} {\bar \psi}\wedge \gamma^{abc} D\psi \wedge e^d \; .
\end{equation}
To see that this is the case, let us first discuss for
a generic superalgebra the problem of local invariance of a
a class of lagrangians depending on its gauge fields and
their curvatures. This includes as a particular
case those depending only on the curvatures (as \eqref{maxcur}) .

\section{Geometry of local invariance and the field equations}

Let us introduce here a geometric procedure to discuss the
local invariance of a class of lagrangians.
Let $H$ be a form that is a combination of exterior products of
the gauge forms $A^i$ associated with a generic Lie superalgebra
and of their curvatures, as defined by the Cartan structure equations
$F^i= dA^i+\frac{1}{2} C_{jk}{}^i A^j\wedge A_k$. Thus, in
general, $H=H(A,F)$.
Let us now introduce two inner derivations $i_{F^i}$ and $i_{A^i}$ of
degree $-2$ and $-1$, respectively, defined by $i_{F^i} A^j=0$, $i_{F^i}F^j=
\delta^j_i$ and $i_{A^i}F^j=0$, $i_{A^i}A^j=\delta^j_i$.
Since $dF^i=C^i{}_{jk} F^j\wedge A^k$, the exterior
differential $d$ may be expressed as
\begin{equation}
\label{dFA}
   d= C^i{}_{jk} F^j\wedge A^k i_{F^i} - \frac{1}{2}C^i{}_{jk} A^j\wedge A^k i_{A^i} +
   F^i i_{A^i}\ .
\end{equation}
Then, the commutator $\left[d,i_{F^i}\right]$ is given by
\begin{equation}
\label{commu}
   d i_{F^i} - i_{F^i} d = - C^j{}_{ik}  A^k i_{F^j}- i_{A^i}\ .
\end{equation}

Now, let $B$ be a lagrangian form $B=B(A,F)$ which is a potential
form of $H$, $H=dB$. Then, it follows that the $A^i$ field equation
is given simply by $i_{F^i}H=0$.
Indeed,  let us first compute the variation of $\int_M  B$, where
$M$ stands for Minkowsli space:
\begin{eqnarray}
\label{deltaB}
  \delta \int_M  B & = & \int_M
  \left\{ \delta A^i\wedge  i_{A^i} B + \delta
  F^i \wedge i_{F^i}B\right\}\nonumber\\
  &= &  \int_M   \left\{ \delta A^i \wedge i_{A^i} B +
  \left(d \delta A^i + C^i{}_{jk} \delta A^j\wedge A^k\right) \wedge
  i_{F^i} B \right\}\nonumber\\
  & =&  \int_M  \delta A^i \wedge \left\{ i_{A^i}B
  + d i_{F^i} B - C^l{}_{ji} A^j \wedge i_{F^l} B\right\} \nonumber \\
  &=& \int_M \delta A^i\wedge i_{F^i}H
    \ ,
\end{eqnarray}
where we have integrated by parts the second term in the
second equality above and used eq.~\eqref{commu} for $di_{F^i} B$
in the third one. Thus, the $A^i$ field equation is
\begin{equation}
\label{Aeqom}
i_{F^i}H\equiv E_i=0  \ .
\end{equation}
We note that the differential $d(i_{F^i}H)$ of the $l.h.s$
of the $A^i$ field equation three-form is, since $dH\equiv 0$,
\begin{equation}
\label{diFH}
    d(i_{F^i}H) = - C^j{}_{ik}  A^k \, i_{F^j}H- i_{A^i} H\ ,
\end{equation}
a condition that will be relevant for the local invariance below.\\

Let us now assume that the three-form  $i_{A^i}H$ on $M$ adopts the expression
 $i_{A^i}H = X^j_i \wedge i_{F^j}H$ for some one-forms $X^i_j$.
This means that the $i_{A^i}H$ vanish when the $i_{F^j}H$ do {\it i.e.},
that they vanish on shell (or are zero).
Then, the following lemma holds:\\

\noindent
{\bf Lemma}.
{\it  Let $i_{A^j}H$ vanish on shell or be zero. Then, $i_{A^j}H$ has the
form $i_{A^j}H = X^i_j \wedge i_{F^i}H$. Let us assume that
$X^i_j \wedge i_{F^i}H \neq 0$. Then, the
action is invariant under a local symmetry $\delta A^i$
of the form}
\begin{equation}
\label{noether}
\delta A^i=\delta_{gauge}A^i+\delta' A^i=
d\alpha^i- C^i{}_{jk} \alpha^j A^k +\delta' A^i \equiv D\alpha^i+\delta' A^i \ ,
\end{equation}
where
\begin{equation}
\label{deltapX}
    \delta' A^i= - X^i_j \alpha^j \
\end{equation}
and the sum is extended to the indices $j$ that make $i_{A^j}H \neq 0.$
\begin{proof}
The extra piece $\delta'A^i$ is needed for $\delta A^i$ in \eqref{noether}
to be a symmetry when  $i_{A^j}H\neq 0$ (this will be the case for
the lagrangian $B$ in eq.~\eqref{maxcur}, because $H=dB$, being a differential of
curvatures, will not be given in terms of curvatures only).
Indeed, as an arbitrary variation of the
action has the form $\delta \int_M B= \int_M  \delta
A^i \wedge i_{F^i}H$, the specific  $\delta
A^i$ in \eqref{noether} leads to
\begin{eqnarray}
\label{expl}
\delta \int_M B &=&
\int_M  \left( d\alpha^i- C^i{}_{jk} \alpha^j A^k
+\delta' A^i \right)\wedge i_{F^i}H
\nonumber \\
&=& -\int_M  \alpha^i \left( d\left(i_{F^i}H\right) +
C^j{}_{ik} A^k \wedge i_{F^j}H\right) + \int_M  \delta'A^i\wedge i_{F^i}H \nonumber\\
&=& \int_M  \left(\alpha^i i_{A^i}H+ \delta' A^i\wedge i_{F^i}H \right) \ ,
\end{eqnarray}
where \eqref{diFH} has been used in the last equality.
We see that the last line of \eqref{expl} vanishes for
$\delta' A^i$ given by \eqref{deltapX}.
\end{proof}
\noindent
In our case, it is at this point where the factor $k$ in \eqref{maxcur}
is fixed so that $B$ becomes the supergravity langrangian.
Note that no on-sell condition has been used; only
the expression of the three-forms $E_i\equiv i_{F^i}H$ that
determine the field equations  ($E_i=0$) enter
in the four-form $\delta A^i\wedge i_{F^i}H$.\\

The above procedure is reminiscent of the construction of
bosonic Chern-Simons (C-S) lagrangians in odd dimensions,
where $H$ is a (symmetric, gauge invariant and closed) even
polynomial in the curvatures and $B$ in  $H=dB$ is the C-S form.
In this C-S case, $i_{A^i}H$ is identically
zero ($H \neq H(A)$) and $\delta \int_M B=0$ without
any $\delta' A^i$ term so that  $\delta A^i=\delta_{gauge} A^i$ is a
genuine gauge transformation and, as we know, $\delta_{gauge} B$ a total derivative.
To derive this in the above context, let $B$ be
a C-S lagrangian. If the two total differentials that were
discarded in eqs.~\eqref{deltaB} and \eqref{expl} are
restored and \eqref{commu} is used, then we obtain
\begin{equation}
\label{deltaC}
     \delta_{gauge} B = d \left(\alpha^i i_{A^i} B \right)\ .
\end{equation}
Let us now check that this formula reproduces the familiar
expression for the gauge variation of a C-S lagrangian form
(see {\it e.g.} \cite{CUP}). Let $A= A^i T_i$, $F= F^i T_i$ and
$\alpha=\alpha^i T_i$, where
$[T_i,T_j]= C^k{}_{ij}T_k$ . Then, the C-S forms may be constructed
as potentials of the closed (Chern) $2l$-forms
\begin{equation}\label{Hl}
    H_{l} = Tr (F\wedge \overset{l}{\dots} \wedge F)\ ,\ dB_l = H_l \
\end{equation}
(we ignore an $l$-dependent factor). Explicitly,
\begin{equation}
\label{explB}
 B_l= l\int^1_0 Tr(A\wedge F_t\wedge \overset{l-1}{\dots} \wedge F_t) \,
\delta t
\end{equation}
where $F_t= t F +(t^2-t) A\wedge A= t dA+ t^2 A\wedge A$. Using \eqref{explB}
and  \eqref{deltaC}, the following formula for the gauge variation
is obtained:
\begin{eqnarray}
\label{Bvar}
\delta B_l &=& d\Big[ l \int_0^1 Tr(\alpha F_t\wedge \overset{l-1}{\dots}
\wedge  F_t \nonumber \\
&-&(t^2-t) A \wedge \sum^{l-2}_{k=0} F_t\wedge  \overset{k}{\dots} F_t
\wedge \left[ \alpha,A\right] \wedge F_t\wedge  \overset{l-2-k}{\dots}\wedge
F_t)\delta t\Big] \, .
\end{eqnarray}

For instance, for $l=2$ ($D$=3),
\begin{eqnarray}
\label{Bvarex}
 \delta B_2 &=& d\,\large[ 2 Tr(\alpha\, F) \int_0^1 t\, \delta t + 6 Tr
 (\alpha A\wedge A) \int_0^1 (t^2-t) \delta t \large] \nonumber\\
 &=& d\,[ Tr\, \alpha \, (F - A\wedge A)] = d\, [Tr(\alpha\, dA)]\ ,
\end{eqnarray}
Similarly, for $l=3$, $D=5$, eq.~\eqref{Bvar} gives
\begin{eqnarray}
\label{l3D51}
\delta B_3 &=& d\, ( \int_0^1 Tr [ \alpha \,F_t\wedge F_t
 \nonumber \\
& - &  (t^2-t) ( A\wedge [\alpha, A]\wedge F_t + A \wedge F_t
\wedge  [\alpha, A] ) ] \, \delta t  ) \ .
\end{eqnarray}
Inserting now the expression of $F_t$ and evaluating the integrals,
one obtains:
\begin{equation}
\label{l3D52}
\delta B_3 = d\, Tr (   \alpha\,  d [  A\wedge dA + \frac{1}{2}
A\wedge A\wedge A ] )\ .
\end{equation}
Both $\delta B_2$ and $\delta B_3$ reproduce
the well known expressions for the variation of the C-S three- and
five-forms under the infinitesimal gauge function $\alpha$.

\section{Pure supergravity from $s\mathcal{M}$}

 Let us apply the above to the superMaxwell algebra
case. First, we compute the differential of $B$ in \eqref{maxcur}.
$H=dB$ is given by
\begin{eqnarray}
\label{dBK}
H &=& 2\epsilon_{abcd} R^{ab} \wedge T^c \wedge e^d
+\left( 1-\frac{k}{8}\right) \epsilon_{abcd} R^{ab} \wedge {\bar \Xi}
\gamma^{cd} \wedge \psi \nonumber\\
& & - \left( 1-\frac{k}{8}\right)\epsilon_{abcd} R^{ab}\wedge  {\bar \xi}
\gamma^{cd} \wedge \Psi + \frac{k}{2} {\bar \Psi} \wedge e_a\gamma^a
\gamma_5 \wedge \Psi \nonumber\\
& & - \frac{k}{2} {\bar \psi} \wedge T_a\gamma^a
\gamma_5 \wedge \Psi\ .
\end{eqnarray}
Now, we observe that $H \neq H(F^{ab})$ and that, when $k$=8,
the $\Xi^\alpha$ dependence is also absent
from $H$; in fact, $H\neq H(\xi^\alpha, \Xi^\alpha, f^{ab}, F^{ab})$.
Thus, both the $\xi^\alpha$ and the $f_{ab}$ field equations
are trivial for $k$=8. This implies that the fields
$\xi^\alpha$ and $f^{ab}$ are not relevant in the action, since they
have to appear in the lagrangian as total derivatives. For the same
value of $k$, the $\psi$ dependence of $H$ is reduced to the last
term in \eqref{dBK} so that
\begin{equation}
\label{ipsiH4}
   i_{{\bar \psi}^\alpha}H = -4 ({\bar\Psi}\wedge T_a \gamma^a \gamma_5 )_\alpha \ .
\end{equation}
Morevover, the $\omega_{ab}$ field equation $E_{ab}=0$,
where $E_{ab} \equiv i_{R^{ab}}H$, and the $T^a=0$ equation imply
each other since the vielbein is invertible. Since \eqref{ipsiH4}
 is, through $T^a$, related to the equation of motion of $\omega^{ab}$,
 this means that in $i_{\psi^\alpha} H = X^i_\alpha \wedge E_i$ the
 only non-vanishing $X^i_\alpha$ corresponds to $i=(ab)$ {\it i.e.},
 to $X^{ab}_\alpha$. Thus (see \eqref{noether}) only a certain
 $\delta{'}_\epsilon \omega^{ab}$ is needed for local supersymmetry
 invariance since for $\delta_\epsilon \psi$ and  $\delta_\epsilon e$
 no $\delta_\epsilon'$ piece appears.

Since for $k=8$ the extra one-form fields $\xi^\alpha$ and $f^{ab}$
are not relevant in the action, it is sufficient to consider
the procedure in Sec.~4 for the one-form fields
$A^i=(\omega_{ab},\psi^\alpha ,e_a)$ and their
curvatures $F^i=(R_{ab},\Psi^\alpha, T_a)$.
From eqs.~\eqref{maxwelld4} and \eqref{DFsuperM} it is
easy to see that, when acting on the relevant variables
($H=H(E^a,T^a,\psi^\alpha, \Psi^\alpha, R^{ab})$),  the Lorentz
covariant exterior differential is given by
\begin{eqnarray}
\label{Dourcase}
  D &=& \left( \frac{1}{2} {\bar \psi} \gamma^a\wedge \psi +T_a\right)
  i_{e_a} + \Psi^\alpha i_{\psi^\alpha} \nonumber\\
  & &  + (R_{ab}\wedge e^b -{\bar \Psi} \gamma_a \wedge \psi) i_{T_a}
  + \frac{1}{4}\left( R_{ab} \gamma^{ab}\right)^\alpha{}_\beta
  \wedge \psi^\beta i_{\Psi^\alpha} \ .
\end{eqnarray}
It follows from this expression that
({\it cf.} eq.~\eqref{commu})
\begin{equation}
\label{iDDi}
    D i_{\Psi^\alpha} -i_{\Psi^\alpha} D =({\bar \psi}\gamma^a)_\alpha i_{T^a}
    -i_{\psi^\alpha}\ .
\end{equation}
   So writing respectively
 $E_\alpha\equiv i_{\Psi^\alpha}H=0$, $E_a \equiv i_{T^a}H=0$
 for the $\psi^\alpha$ and $e^a$ field equations, the Lorentz covariant
 exterior differential of the R-S three-form $E_\alpha$ in
 the R-S equation satisfies
 \begin{equation}
 \label{DRarita}
    DE_\alpha - (\bar{\psi} \gamma^a)_\alpha E_a =-i_{\psi^\alpha} H\ .
\end{equation}

Then, under the local supersymmetry variations
\begin{eqnarray}
\label{deltaepsilon}
\delta_\epsilon  \psi^\alpha = D\epsilon^\alpha \quad ,\quad
\delta_\epsilon  e^a= {\bar \epsilon} \gamma_a\psi \ ,
\end{eqnarray}
(since $\delta_\epsilon' \psi=0,\,\delta_\epsilon' e=0$)
plus a certain non-zero variation $ \delta_\epsilon' \omega^{ab}$,
the action is invariant:
\begin{eqnarray}
\label{deltaI}
\delta_\epsilon \int_M  B & =& \int_M  \left( \delta_\epsilon
\psi^\alpha \wedge E_\alpha + \delta_\epsilon e^a \wedge E_a+ \delta_\epsilon' \omega^{ab}
\wedge E_{ab}
\right)\nonumber\\
&=& \int_M  \left(D\epsilon^\alpha \wedge
E_\alpha + \bar{\epsilon} \gamma^{a} \psi
\wedge E_a + \delta_\epsilon' \omega^{ab}
\wedge E_{ab} \right) \nonumber\\
&=& \int_M  \left( -\epsilon^\alpha DE_\alpha +
\bar{\epsilon} \gamma^{a} \psi
\wedge E_a + \delta_\epsilon' \omega^{ab}
\wedge E_{ab} \right) \ .
\end{eqnarray}
\noindent
Using now \eqref{DRarita} in \eqref{deltaI} leads to
\begin{equation}
\label{deltaIb}
\delta_\epsilon \int_M  B = \int_M  \left( \delta_\epsilon' \omega^{ab}
\wedge E_{ab} + \epsilon^\alpha i_{\psi^\alpha} H \right) \ .
\end{equation}

Now, there exists on $M$ a set of one-forms
$X_\alpha{}^{ab}$ such that
\begin{equation}
\label{Xab}
i_{\psi^\alpha}H= X_\alpha{}^{ab} \wedge i_{R^{ab}}H \equiv X_\alpha^{ab}\wedge E_{ab}\ ,
\end{equation}
where $i_{\psi^\alpha} H= -4({\bar \Psi} \wedge T_a \gamma^a
\gamma_5)_\alpha$ and $i_{R^{ab}}H = 2\epsilon_{abcd} T^c \wedge
e^d$. A computation shows that the one form $X_\alpha{}^{ab}$
is given by
\begin{equation}\label{Xabp}
   X_\alpha{}^{ab} = -\frac{1}{2}(\epsilon^{abcd} e^g + \epsilon^{bcdg} e^a)\wedge
    ({\bar \Psi}_{cd}\gamma_g \gamma_5)_\alpha - (a\leftrightarrow
    b)\ ,
\end{equation}
where ${\bar \Psi} = {\bar \Psi}_{cd}\, e^c\wedge e^d $. Then, using \eqref{Xab} in
\eqref{deltaIb}, we find that there is local supersymmetry for
\begin{equation}
\label{varomega}
    \delta_\epsilon' \omega^{ab} = - \epsilon^\alpha X_\alpha{}^{ab}
\end{equation}
with $X_\alpha{}^{ab}$ given by \eqref{Xabp},
which is seen to coincide with the well known local supersymmetry variation
of $\omega$. Thus, the lagrangian \eqref{maxcur} for $k$=8 is local
sypersymmetry invariant. As for the local Lorentz and translation
variations, the same general pattern of Sec.~4
applies. For the Lorentz variations, $i_{\omega^{ab}}H = 0$
since $H$ in \eqref{dBK} does not depend on $\omega_{ab}$. Thus
$X^i_{ab}=0$ for all values of $i$, and there is no $\delta'$
(the action is directly Lorentz invariant). For the local
translations, however, $i_{e^a}H\neq 0$. In fact, besides the pieces
containing $T_a$, and hence related to $E_{ab}$, there is the piece
$4 \bar{\Psi}\gamma_a \gamma_5 \wedge \Psi$, which can be shown
to be related to $E_\alpha$ ($E_\alpha=0$ being the R-S equation)
by $X_a^\alpha = \Psi^\alpha_{ab} e^b$. Therefore,
$X_a^\alpha$ and $X_a^{bc}$ are different from zero
so that, besides the piece $\delta{'}_\epsilon \omega^{ab}$ in the
variation, there is also $\delta{'}_{t^a} \psi^\alpha = - \Psi^\alpha_{ab} t^b$,
$t^a$ being the local translation.

 We may finally show that the lagrangian \eqref{maxcur} for $k$=8
is that of pure $D$=4 supergravity. It may be
rewritten in the form
\begin{eqnarray}
\label{lsugra2}
B &=& \epsilon_{abcd} R^{ab}\wedge e^c\wedge e^d + 4 \bar{\psi} \wedge e_a \gamma^a
\gamma_5\wedge \Psi\nonumber\\
& & + d\left( \epsilon_{abcd}  R^{ab}\wedge f^{cd} + 8 \bar{\xi}
\gamma_5 \Psi  \right) \ ,
\end{eqnarray}
which is the $D$=4 simple supergravity lagrangian \cite{FNF76,DZ76}, \cite{PvN-81}
(eq.~\eqref{Actiongrav} plus eq.~\eqref{acRS}) but for the total derivative
in the second line.

\section{Final comments}
We have shown that the first-order Lagrangian four-form
of $D$=4 minimal ($N$=1) supergravity can be written out of bilinears of the
curvatures of the gauge fields associated with the
minimal Maxwell superalgebra of \cite{BGKL10},
thus generalizing the results for gravity in \cite{AKL11} to the supergravity case.
The action is the sum of two terms in the $s\mathcal{M}$ curvatures, and for
a certain relative factor the extra gauge field forms not contained in the supergravity
supermultiplet enter in the action inside a total derivative. For this relative
factor, the sum gives the action of minimal $D$=4 supergravity
as shown by eq.~\eqref{lsugra2}.

This provides one more example of how new geometrical
aspects of a theory may be exhibited by formulating it on the enlarged
superspaces associated to larger algebras, even if the
additional fields in the enlarged superspace
variables/fields correspondence (see \cite{CAIP00, deAI-01,BAIPV04})
do not have a dynamical character. In the present case, the enlarged
superspace would be determined by the supergroup coset {\it sM/L} and
would contain, besides the four-dimensional Minkowski spacetime,
6 bosonic tensorial variables and the 4+4 fermionic ones of the two
Majorana spinors.

Going beyond $D$=4 presents difficulties. An obvious one is the fact that
in odd dimensions there is no way of writing a Lagrangian $D$-form out
of curvature two-forms. This would seem to indicate that in odd dimensions the appropriate
point of view is to look for Lagrangians the differentials of which
are written solely in terms of the curvatures. This is, of course, the case of
actions of the Chern-Simons type. For instance, for $D$=3, the
$(p+q)$ supergravities \cite{AP-89} are C-S theories \cite{AzIz-11}
for an expansion of $osp(p+q|2,\mathbb{R})$ .
Another difficulty, also present in $D$=4, $N>1$,
is the existence in some cases  of Lagrange multiplier
 zero-forms in the first-order actions, which cannot be interpreted in terms
of one-form fields for Lie superalgebras. In higher dimensions there are also
forms of order higher than one, but these could, in principle, be given a
group theoretical interpretation, as done for $D=11$ supergravity
(see \cite{DF82,BAIPV04}).
Another problem of higher even-dimensional supergravities in our scheme
is that,  in the present $D$=4 bosonic case, the extra field
$f^{ab}$ has trivial equations of motion since there is a
single $F^{ab}$ curvature in the lagrangian (see eq.~\eqref{maxcur}),
which would not be the case for $D>$4.

It would be interesting to look further at the role
of the relative weight of the two terms in the basic supergravity lagrangian,
as well as the effect of other possible lagrangian terms in
\eqref{maxcur} which, in the case of simple gravity, lead
to a generalized cosmological term \cite{AKL11}. Another
case worth studying in an analogous approach would be that of
$D$=4, $N=1$ $AdS$ supergravity. \\

\noindent
{\it Note added.} For further work on Maxwell superalgebras see
\cite{CR:14} \\

\noindent
{\bf Acknowledgements}. This work has been partially supported by
the research grant CONSOLIDER CPAN-CSD2007-00042.

\end{document}